# Cloud Template, a Big Data Solution

Mehdi Bahrami

Electronic Engineering and Computer Science Department

University of California, Merced, USA

MBahrami@UCMerced.edu

*Abstract.* **Today cloud computing has become as a new concept for hosting and delivering different services over the Internet for big data solutions. Cloud computing is attractive to different business owners of both small and enterprise as it eliminates the requirement for users to plan ahead for provisioning, and allows enterprises to start from the small and increase resources only when there is a rise in service demand. Despite the fact that cloud computing offers huge opportunities to the IT industry, the development of cloud computing technology is currently has several issues. This study presents an idea for introducing cloud templates which will be used for analyzing, designing, developing and implementing cloud computing systems. We'll present a template based design for cloud computing systems, highlighting its key concepts, architectural principles and state-of-the-art implementation, as well as research challenges and future work requirements. The aim of this idea is to provide a better understanding of the design challenges of cloud computing and identify important research directions in this big data increasingly important area. We'll describe a series of studies by which we and other researchers have assessed the effectiveness of these techniques in practical situations. Finally, in this study we will show how this idea could be implemented in a practical and useful way in industry.**

* Corresponding Author:
Mehdi Bahrami,
Electronic Engineering and Computer Science Department,
University of California, Merced,
5200 Lake Road, Merced, CA 95343, USA
Email: MBahrami@UCMerced.edu

## 1. Motivation

This idea is based on concept of natural cloud [2]. Different weather is based on different cloud on the sky. So, we can use different cloud template as based for designing a cloud computing [8, 9, 10] systems [12] with different character [14, 17].

First of all, this idea back to concept of various types of clouds on the sky. As shown in Figure 1, we have different characters of the cloud and each character has different style and makes different weather. It means different types of a cloud are useful for different atmospheres and geographies. As first step, we'll consider this idea which is based on natural cloud then we'll back to cloud template architecture.

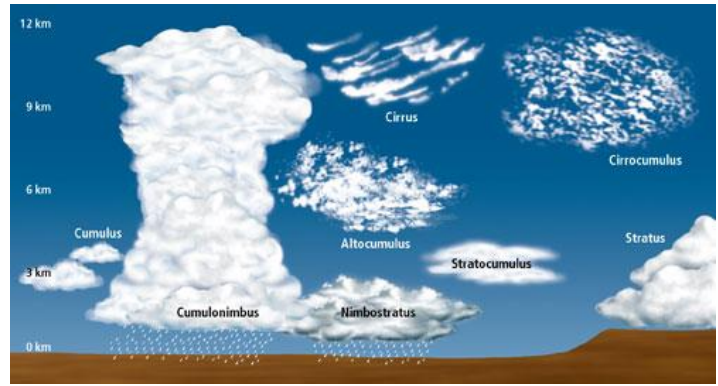

Figure 1. Various types of clouds [2]

## 2. Various types of Clouds:

Various types of clouds are available [2] on the sky and each of them has different purpose:

### 2.1. High-level Clouds:

High-level clouds occur above about 20,000 feet and are given the prefix "cirro-". Due to cold tropospheric temperatures at these levels, the clouds primarily are composed of ice crystals, and often appear thin, streaky, and white.

### 2.2. Mid-level Clouds:

The bases of clouds in the middle level of the troposphere, given the prefix "alto-", appear between 6,500 and 20,000 feet. Depending on the altitude, time of year, and vertical temperature structure of the troposphere, these clouds may be composed of liquid water droplets, ice crystals, or a combination of the two, including super cooled droplets.

### 2.3. Low-level Clouds:

Low-level clouds are not given a prefix, although their names are derived from "strato-" or "cumulo-", depending on their characteristics. Low clouds occur below 6500 feet, and normally consist of liquid water droplets or even super cooled droplets, except during cold winter storms when ice crystals (and snow) comprise much of the clouds.

## 3. Proposed Idea, Cloud Template:

As mentioned in the last section each type of cloud has one or more character [11]. It means, we can use this feature [16,17] in computing systems. We need different type [18, 19] of computing in a cloud computing system then we can use a cloud based on selected template. This proposed idea which we called "Cloud Templates", shown in Figure 2 and shows how we can design a cloud computing system by using a cloud template.

As a first step for proposed idea, we should choose a preferred cloud [20, 22] template [26] which is designed in high level [17, 27] abstract. Then we will have a customized [21, 23, 29] cloud architecture [7, 8] based on the selected cloud template. As a result, we will have different cloud computing systems for different applications [8, 13] with collaboration between each of them [15, 30].

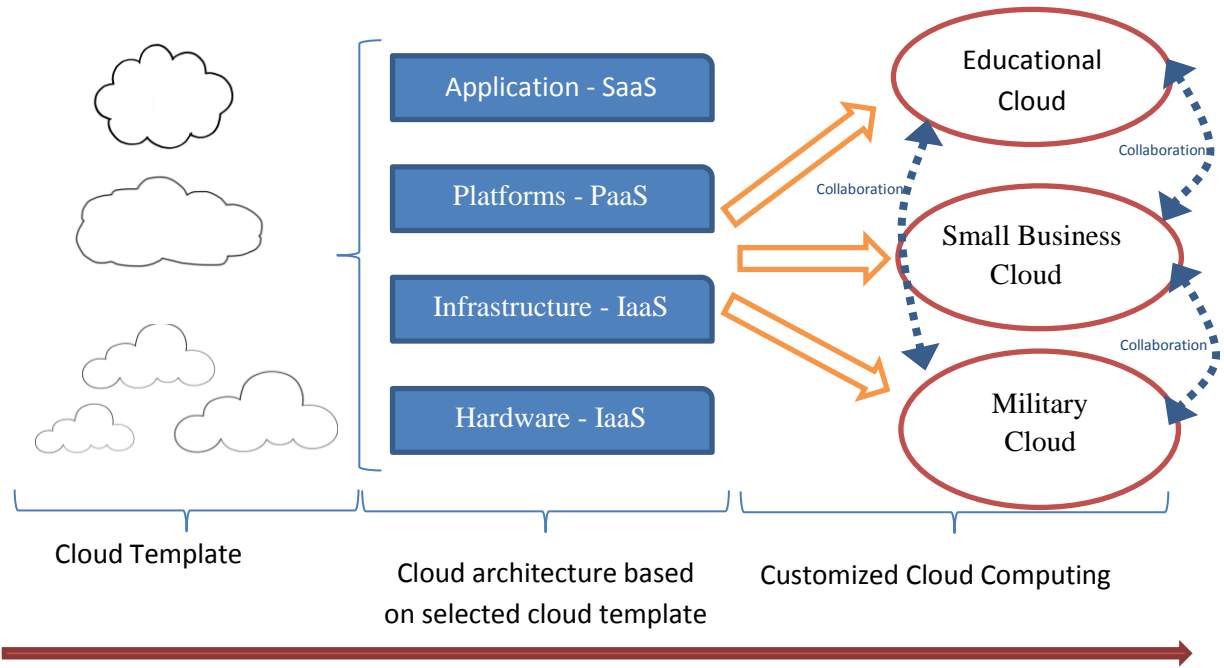

**Figure 2.** Template based Design

## 4. Why we need different cloud templates:

As mentioned in [1][3],[4],[5] and [7], we have several issues in designing a cloud computing system and we need a high-level design to overcome to potential issues. We can use cloud templates to solve or improve these issues:

1- **Complexity:** Cloud computing systems often are complex. So, by introducing different level of abstract in high level design and implementation of a cloud by a template, we can overcome to complexity.
2- **Flexibility in cloud architecture:** This proposed idea helps us to customize a cloud computing architecture based on selected template.
3- **Different templates for different applications:** Introduce different type of cloud computing systems based on different application.
4- **Different templates for different business types:** Small business owners need small architecture. A cloud template helps a small business owner to have their own cloud, without considering detail of the cloud. However other business, such as enterprise, they should using other cloud template for more detail and using this idea for different applications.
5- **High level design:** When we designing a cloud template with a security plan [24, 27], stability plan [27], reliability plan [24, 27] in a high-level design [28], then this template could be using in the lower level with more reliable, secure and stable.
6- **Define cloud collaboration in high level design:** When we designing a template and customized it for future collaboration [21, 29, 30], then we'll have simple way for collaborating in the cloud.

## 5. Future Work and conclusion:

This is a new idea and it's requiring more study as we listed below:

1- Implementation some tools for developing and designing a cloud template and features.
2- How we can have different abstract level for cloud computing systems based on each template.
3- How the proposed idea could customize the cloud computing layer architecture (SaaS, PaaS, IaaS, IaaS (Hardware) ).
4- How a cloud template could help us in a cloud computing collaboration between different clouds.
5- Connection between cloud template to Public, Private and Hybrid clouds.

With the emergence of cloud computing system as a computer science paradigm in which computing is done exclusively on resources leased only when needed from big data centers, scientists are faced with a new platform option. However, the initial target often cloud computing system does not match the characteristics of the scientific computing workloads, also often scientists are require customize their cloud based on requirement. In this paper we introduced an idea in high level design for overcoming to cloud computing issues. Our main finding is that the cloud computing systems are requiring a revolution such as using cloud template for different purpose on a cloud.